\documentclass[twocolumn]{aastex62}

\newcommand{\cnjwl}{$cn_{\rm JWL}$}
\newcommand{\dcnjwl}{$\Delta_{\rm r}cn_{\rm JWL}$}

\newcommand{\cnjwlcor}{$cn_{\rm JWL,cor}$}

\newcommand{\cacn}{Ca-CN}
\newcommand{\cnw}{CN-w}
\newcommand{\cns}{CN-s}
\newcommand{\vvhb}{$V - V_{\rm HB}$}

\newcommand{\vhb}{$V_{\rm HB}$}

\newcommand{\str}{Str\"omgren}

\newcommand{\nrgb}{$n$(\cnw):$n$(\cns)}
\newcommand{\nfgsg}{$n$(FG):$n$(SG)}

\newcommand{\ebv}{$E(B-V)$}

\newcommand{\hst}{{\it HST}}

\newcommand{\sdss}{SDSS}

\newcommand{\cnwave}{$\lambda$3883}
\newcommand{\chwave}{$\lambda$4250}
\newcommand{\scn}{$S$(3839)}
\newcommand{\ds}{$\delta S$(3839)}

\newcommand{\dug}{$\Delta^\prime_{u-g}$}
\newcommand{\dc}{$\Delta_{\rm F275W,F814W}$}
\newcommand{\dtrio}{$\Delta_{\rm C~F275W,F336W,F438W}$}
\newcommand{\trio}{$C_{\rm F275W,F336W,F438W}$}

\newcommand{\cubi}{$C_{UBI}$}
\newcommand{\dcubi}{$\Delta_{\rm r}C_{UBI}$}

\newcommand{\sres}{$\lambda_{\rm C}/\Delta\lambda$}
\newcommand{\dw}{$\Delta\lambda$}

\shorttitle{\cnjwl\ versus \cubi}
\shortauthors{Lee}

\begin{document}

\title{MULTIPLE STELLAR POPULATIONS OF GLOBULAR CLUSTERS FROM 
HOMOGENEOUS \cacn\ PHOTOMETRY. 
IV. TOWARD PRECISION POPULATIONAL TAGGING.
\footnote{Based on observations made with the Cerro Tololo Inter-American Observatory (CTIO) 1 m telescope, which is operated by the SMARTS consortium, and the Kitt Peak National Observatory (KPNO) 0.9 m telescope, which is operated by WIYN Inc. on behalf of a Consortium of partner Universities and Organizations.}}

\author[0000-0002-2122-3030]{Jae-Woo Lee}
\affiliation{Department of Physics and Astronomy, Sejong University\\
209 Neungdong-ro, Gwangjin-Gu, Seoul, 05006, Korea\\
jaewoolee@sejong.ac.kr, jaewoolee@sejong.edu}

\begin{abstract}
Apparently similar but multifaceted photometric systems are currently being used to investigate the multiple stellar populations in globular clusters, without the concrete general agreement on the definition of the multiple populations. In recent years, an attractive idea of utilization of the widely used $UBI$ photometry, \cubi, for the populational tagging of the giant stars in globular clusters has been emerged. We perform a critical analysis of the \cnjwl\ and the \cubi\ indices, finding that the populational tagging from the \cubi\ index may not be reliable, due to the inherited trait of the broad-band photometry. As a consequence, the populational number ratios and the cumulative radial distributions from the \cubi\ index can be easily in error. The results for M3, which shows a very strong radial gradient in the populational number ratio, highlights the strengths of our \cnjwl index: both the HST imaging and the ground-based spectroscopy failed to grasp the correct picture, that can be easily achieved with our \cnjwl\ index with small aperture ground-based telescopes, due to the small field of view or crowdedness in the central part of the cluster.
\end{abstract}


\section{INTRODUCTION}
The most outstanding achievement in stellar astrophysics during the past decade would be the discovery of the multiple populations (MPs) in globular clusters (GCs). By and large, the decades-long unsolved puzzle of the ubiquitous nature of the CN bimodality and the Na-O anticorrelations seen in GCs can be understood in the context of the MPs in normal GCs, although the detailed scenarios are still not certain.

Populational number ratios and the cumulative radial distributions can provide the core information of the formation and the evolution of the GCs with MPs. For example, the so-called mass-budget problem seen in Galactic GCs is one of the most quizzical problems, and it should be understood on the basis of the correct populational tagging in individual GCs \citep[see, e.g.,][]{renzini15}. Also, the cumulative radial distributions of the MPs in GCs are closely linked to the dynamical evolutionary state, and subtle incorrect populational tagging can lead to an utterly different conclusion. Therefore, securing the correct populational tagging of individual stars in GCs from the large field of view (FOV) is the first step toward better understanding of the GC formation and evolution.

The lack of consensus on the definition of the MPs in GCs is a pending issue of great importance. Seemingly similar but fundamentally different bases of the populational tagging can misguide us, as we will show later. Basically, the MPs in the high-resolution spectroscopic study are often defined by the oxygen and sodium abundances \citep[see][]{carretta09}. Besides the difficulty in the oxygen abundance measurements, there are other aspects that make the spectroscopic abundance analysis vulnerable \citep{lee10,lee16}. Also importantly, the MPs assigned from the Na--O plane are somewhat arbitrary, since the populational tagging from the apparently continuous Na--O anticorrelations appears to be somewhat factitious. On the other hand, the populational tagging based on the nitrogen abundances is rather straightforward in some metallicity ranges owing to the presence of the discrete bimodal nitrogen abundance distribution in the GC stars.

There are at least a few photometric systems currently being used in the study of the MPs in GCs, but, very unfortunately, the definition of MPs in GCs from various photometric studies can differ one from the other \citep[see, e.g.,][]{lardo11,lardo17,jwlnat,jwln1851,lee17,lee18,milone12,milone17}. The MPs study from broadband photometry tends to rely on the photometric measurements of absorption strengths of some very strong molecular bands, including OH, NH, CN, and CH. In particular, \citet{milone12} devised an interesting color index, $C_{UBI} = (U-B) - (B-I)$, using the widely used Johnson--Kron/Cousins photometric system to study the MPs in GCs \citep[see also][]{monelli13}. It would be great if \cubi\ can perform accurate populational tagging for stars in GCs, since there is a huge amount of archival photometric data available. Also, being based on the broad-band system, the integration time for the \cubi\ index can be significantly reduced compared to our \cnjwl\  (= JWL39 $-$ $Ca_{\rm new}$) index \citep[see][]{lee17,lee18}.

In this paper, we perform a critical assessment between our \cnjwl\ and the \cubi\ indices. As we already showed in our previous studies \citep{lee17,lee18}, the photometric products from broadband systems can suffer from confusion, and accurate populational tagging can be a demanding task. For example, we showed that the populational tagging of the M5 RGB stars from the \sdss\ photometry by \citet{lardo11} may not be satisfactory and we cautioned about its ability: the \sdss\ system should not be used for the precision populational tagging, similar to what we have shown for the $m1$ [= $(v-b)-(b-y)$] and the $cy$ [= $c1 - (b-y)$ = $(u-v)-(v-b) - (b-y)$; \citealt{yong08}]. In our current work, we decline to discuss the utility of the $m1$ and the $cy$ indices. We elaborately examined that neither the $m1$ nor the $cy$ are good measures of the lighter elemental abundances, and therefore their utilities of the populational tagging for MPs in normal GCs should be very limited \citep[see,][]{lee17,lee18}.

%%%%%%%%%%%%%%%%%%%%%%%%%%%%%%%%%%%%%%%%%%%%%%%%%%%%%%%%%%%%%%
\begin{figure}
\epsscale{1.2}
\figurenum{1}
\plotone{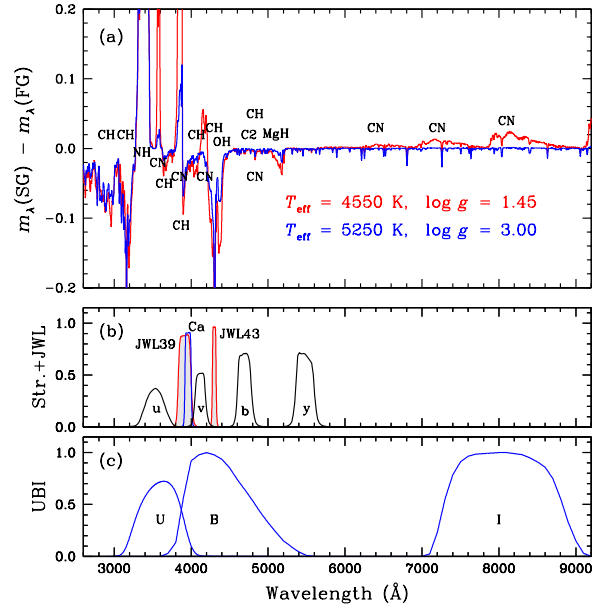}
\caption{
(a) Comparisons of synthetic spectra for RGB stars with intermediate metallicity ([Fe/H] = $-$1.5). The red solid line shows the difference in the monochromatic magnitude between the bright ($V = V_{\rm HB} - 1.5$ mag) FG and the SG of the stars, and the blue solid line shows that for the faint stars ($V = V_{\rm HB} + 1.5$ mag). 
(b) Filter transmission curves of the extended \str\ and the JWL systems. The \str\ $uvby$ are shown with black solid lines, $Ca_{\rm new}$ with the blue line, and $JWL39$ and $JWL43$ with red solid lines. Note that $JWL39$ measures the CN $\lambda$3883 and $JWL43$ measures the CH $\lambda$4300 absorption strengths. 
(c) Filter transmission curves of the UBI system.
}\label{fig:filter}
\end{figure}
%%%%%%%%%%%%%%%%%%%%%%%%%%%%%%%%%%%%%%%%%%%%%%%%%%%%%%%%%%%%%%

%=======================================================================
\begin{deluxetable*}{cccccccccc}
\tablenum{1}
\tablecaption{Integration times for M3 (s)\label{tab:obs}}
\tablewidth{0pc}
\tablehead{
\colhead{} &
\colhead{$y$} & \colhead{$b$} & \colhead{$Ca_{\rm new}$} & \colhead{$JWL39$} & 
\colhead{} &
\colhead{$V$} & \colhead{$B$} & \colhead{$I$} & \colhead{$U$}
}
\startdata
NGC~5272(M3)    & 7690 & 17320 & 52000 & 23500 & & 1380 & 3070 & 1220 & 11350 \\
\enddata 
\end{deluxetable*}
%=======================================================================

\section{PHOTOMETRIC DATA}\label{s:data}
For our current study, we used our own photometry for the extended \str\ and JWL filter systems \citep{lee15}.

The photometric data for M5 and NGC~6752 were collected using the CTIO 1 m, and the results for these two clusters were already published elsewhere \citep{lee17,lee18}. The detailed discussions of the observing procedures and the instrument setup for our CTIO observations can be found in \citet{lee15}. For $UBVRI$ photometry of these two GCs, we used the photometric data kindly supplied by Dr. Stetson (Steton, P.~B.\ 2018, private communication).

The photometric data for M3 were obtained using the KPNO 0.9 m telescopes. In 2017 and 2018, we observed M3 in 20 nights in five separate runs using  the Half Degree Imager (HDI), which is equipped with an e2V 4k $\times$ 4k CCD chip providing a FOV of 30\arcmin $\times$ 30\arcmin. The total integration times for M3 are given in Table~\ref{tab:obs}.

We note that the interstellar reddening values for these three clusters are very small, \ebv\ = 0.01, 0.03, and 0.04 mag for M3, M5, and NGC~6752, respectively \citep{harris96}. Therefore, the differential reddening will not affect our results presented here, and we do not attempt to correct the differential reddening effects. Also, the Galactic latitudes for these GCs are rather high, 79\degr, 47\degr, and $-$26\degr\ for M3, M5, and NGC~6752, respectively, and therefore the contamination from the off-cluster field stars should not be severe in our results.

%=======================================================================
\begin{figure}
\epsscale{1.2}
\figurenum{2}
\plotone{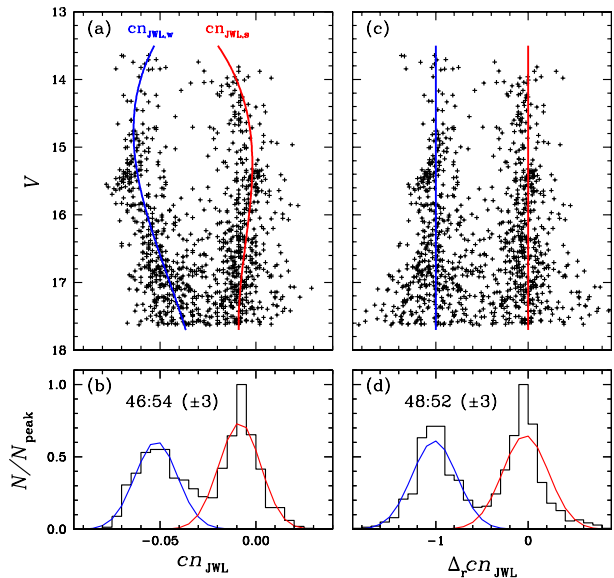}
\caption{
(a) \cnjwl\ CMD of M3 RGB stars. The blue and the red solid lines denote the fiducial sequences for the \cnw\ and the \cns\ RGB populations, respectively. Note that the discrete double RGB populations are noticeable.
(b) Distribution of RGB stars showing two peaks with the number ratio of \nrgb\ = 46:54 ($\pm$ 3).
(c) CMD of M3 RGB stars using the rectified color index, \dcnjwl.
(d) Distribution of RGB stars along \dcnjwl, with the number ratio of \nrgb\ = 48:52 ($\pm$ 3).
}\label{fig:rec}
\end{figure}
%=======================================================================

\begin{deluxetable*}{ccrrrrrrr}
\tablecaption{Spectral resolving powers for selected filters.\label{tab:filter}}
\tablenum{2}
\tablewidth{0pc}
\tablehead{
\multicolumn{1}{c}{} &
\multicolumn{1}{c}{} &
\multicolumn{3}{c}{Johnson} &
\multicolumn{1}{c}{} &
\multicolumn{3}{c}{$Ca$ + JWL} \\
\cline{3-5}\cline{7-9}
\multicolumn{1}{c}{} &
\multicolumn{1}{c}{} &
\multicolumn{1}{c}{$U$} &
\multicolumn{1}{c}{$B$} &
\multicolumn{1}{c}{$I$} &
\multicolumn{1}{c}{} &
\multicolumn{1}{c}{$Ca_{\rm new}$} &
\multicolumn{1}{c}{$JWL39$} &
\multicolumn{1}{c}{\cnjwl\tablenotemark{1}} 
}
\startdata
$\lambda_{\rm c}$ & (nm) 
       & 367 & 436  & 797 &&  395 & 390 &  388 \\
$\Delta\lambda$  & (nm) 
       & 66  & 94   & 149 &&    9 &   18 &    9 \\
$\lambda_{\rm c}/\Delta\lambda$ & 
       & 5.5 & 4.6  & 5.3 && 43.8 & 21.7 &  43.1 \\
\enddata
\tablenotetext{1}{\cnjwl = $JWL39 - Ca_{\rm new}$}
\end{deluxetable*}

\section{PHOTOMETRIC SYSTEMS AND COLOR-MAGNITUDE DIAGRAMS FOR M5}\label{s:filter}
In Table~\ref{tab:filter}, we show the spectral resolving powers for selected filters used in this study (see also Table~1 of \citealt{lee17}). The resolving powers for our $Ca_{\rm new}$ and \cnjwl\ are \sres\ $\approx$ 20 -- 40, while those of the other filters are relatively very low, \sres\ $\lesssim$ 6, owing to very broad bandwidths for such systems, \dw\ $>$ 65 nm.

\subsection{The extended \str\ + JWL system}
Recently, we developed new filters, JWL39 and JWL43, in order to measure the absorption strengths of CN \cnwave\ \AA\ and CH \chwave\ \AA\ molecular bands in cool giants in GCs. In Figure~\ref{fig:filter}, we show the filter bandpasses along with the difference in the monochromatic magnitude between the first generation (FG) and the second generation (SG) of the stars. In our previous work, we introduced a new CN index, \cnjwl, which is an excellent measure of the CN \cnwave\ band and, furthermore, is an excellent nitrogen abundance tracer in giant stars in GCs. As we already showed for M5 and NGC~6752, our \cnjwl\ index is as good as the classical spectroscopic indices, \scn\ or \ds\ \citep[for the definitions of these two indices, see][]{norris81}, and is capable of distinguishing MPs in GCs with great satisfaction \citep{lee17,lee18}.

Being a measure of the CN \cnwave\ \AA\ molecular band feature, our \cnjwl\ index naturally suffers from a weak luminosity effect.In order to remove the luminosity effect, we devised a rectified \cnjwl\ index, defined as
\begin{equation}
\Delta_{\rm r}cn_{\rm JWL} = \frac{cn_{\rm JWL} - cn_{\rm JWL,s}}{cn_{\rm JWL,s}-cn_{\rm JWL,w}},\label{eq1}
\end{equation}
where $cn_{\rm JWL,w}$ and $cn_{\rm JWL,s}$ represent the fiducials of the \cnjwl\ index for the \cnw\ and the \cns\ sequences, respectively, at a given visual magnitude as shown in Figure~\ref{fig:rec}.
Note that the \cnw\ and the \cns\ populations are defined to be groups of stars with smaller and larger \cnjwl\ indices, respectively, at a given visual magnitude. Therefore, the fiducial sequence for the \cnw\ population is located at \dcnjwl\ = $-$1, while that for the \cns\ population is located at \dcnjwl\ = 0. The advantage of employing the  $\Delta_{\rm r}cn_{\rm JWL}$ index is that the populational tagging becomes more straightforward with fixed Gaussians against the $V$ magnitude during our calculations of the expectation values as will be discussed in \S\ref{s:tag}. Also importantly, we can directly compare our \dcnjwl\ index with the rectified \cubi\ index, \dcubi, which will be defined later, in the same scale.

%=======================================================================
\begin{figure}
\epsscale{1.2}
\figurenum{3}
\plotone{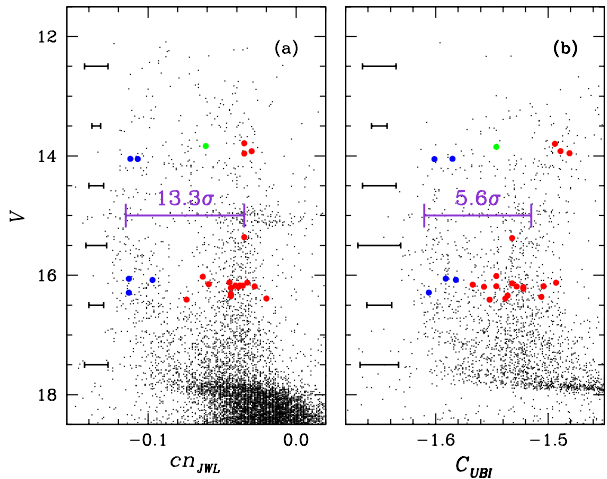}
\caption{
The \cnjwl\ ( = JWL39  $-$ $Ca_{\rm new}$) and \cubi\ [$ = (U-B) - (B-I)$] CMDs of the M5 field with the radial distance of 1\arcmin\ $\leq r \leq$ 10\arcmin. Note that the off-cluster field stars were not removed in the figure, but the contamination from the off-cluster field stars is not expected to be severe. The mean photometric measurement errors in each color index are also shown. The blue, green, and red circles represent the CN-normal, CN-intermediate, and CN-strong RGB stars, respectively, spectroscopically classified by \citet{briley92}. The \cnjwl\ split between the two populations at \vhb\ is 13.3$\times\sigma$(\cnjwl), and the \cubi\ split is 5.6$\times\sigma$(\cubi), where $\sigma$(\cnjwl) and $\sigma$(\cubi) are the photometric measurement uncertainties. As shown, the populational separation in the \cubi\ index is not as clear as that in \cnjwl. Furthermore, the transition from one population to the other is rather continuous in the lower RGB sequence for \cubi. The ambiguous behavior in the \cubi\ index is not due to the measurement uncertainties but due to the intrinsic nature of ambiguity.
}\label{fig:m5cubi}
\end{figure}
%=======================================================================

\subsection{The UBI system}
As we mentioned above, \citet{milone12} devised a color index, $C_{UBI} = (U-B) - (B-I)$, using the widely used Johnson--Kron/Cousins system \citep[see also][]{monelli13}, to study the MPs in GCs. However, as shown in Figure~\ref{fig:filter}, the wavelength coverage for these three filters includes very strong absorption features, such as NH, CN, and CH. The interpretation of the \cubi\ index can be complicated for various reasons. 
(i) The nitrogen and the carbon abundances of RGB stars even in the same stellar population show non-negligible spreads \citep[for example, see][]{cohen02}.
(ii) It is a well known fact that the GC stars exhibit a positive CN--NH correlation and a CN--CH anticorrelation, and therefore the net effect in the \cubi\ index can be intensified or diminished.
(iii) These diatomic molecules show a different degree of the luminosity effects owing to different dissociation energies as shown in Figure~\ref{fig:filter}(a).
(iv) Finally, the wavelength coverage of the \cubi\ index is so large that it is also affected by the continuum of the spectra, which are expected to vary between the MPs in GCs, due to the different elemental abundances.

In Figure~\ref{fig:m5cubi}, we show the \cnjwl\ and \cubi\ color--magnitude diagrams (CMDs) for M5 along with the RGB stars studied by \citet{briley92}. The figure shows that the \cubi\ index is capable of distinguishing  stars with different CN contents from the spectroscopic measurements, but the populational separation in the \cubi\ index is not as clear as that in our \cnjwl\ index. Furthermore, especially in the lower part of the RGB sequence, the transition from the \cubi-blue to the \cubi-red populations becomes continuous. The photometric measurement error of the \cubi\ index cannot explain the confusion in separating MPs in GCs. In the figure, we show the splits between the two RGB populations in the \cnjwl\ and \cubi\ indices at \vhb, and they are at the levels of 13.3$\times\sigma$(\cnjwl) and 5.6$\times\sigma$(\cubi), where $\sigma$(\cnjwl) and $\sigma$(\cubi) are the photometric measurement uncertainties. In both indices, the splits are significantly larger than the photometric measurement errors. Therefore, the confusion in separating MPs in GCs is the intrinsic nature of the \cubi\ index.

As with \dcnjwl, we defined the rectified \cubi\ index, 
\begin{equation}
\Delta_{\rm r}C_{UBI} = \frac{C_{UBI} - C_{UBI,r}}{C_{UBI,r}-C_{UBI,b}},\label{eq2}
\end{equation}
where $C_{UBI,b}$ and $C_{UBI,r}$ represent the the fiducials for the \cubi-blue and \cubi-red sequences, respectively, at a given visual magnitude. As will be shown later, both the \cubi-blue and the \cubi-red sequences are not as distinctive as the \cnw\ and the \cns\ sequences.

%=======================================================================
\begin{figure}
\epsscale{1.2}
\figurenum{4}
\plotone{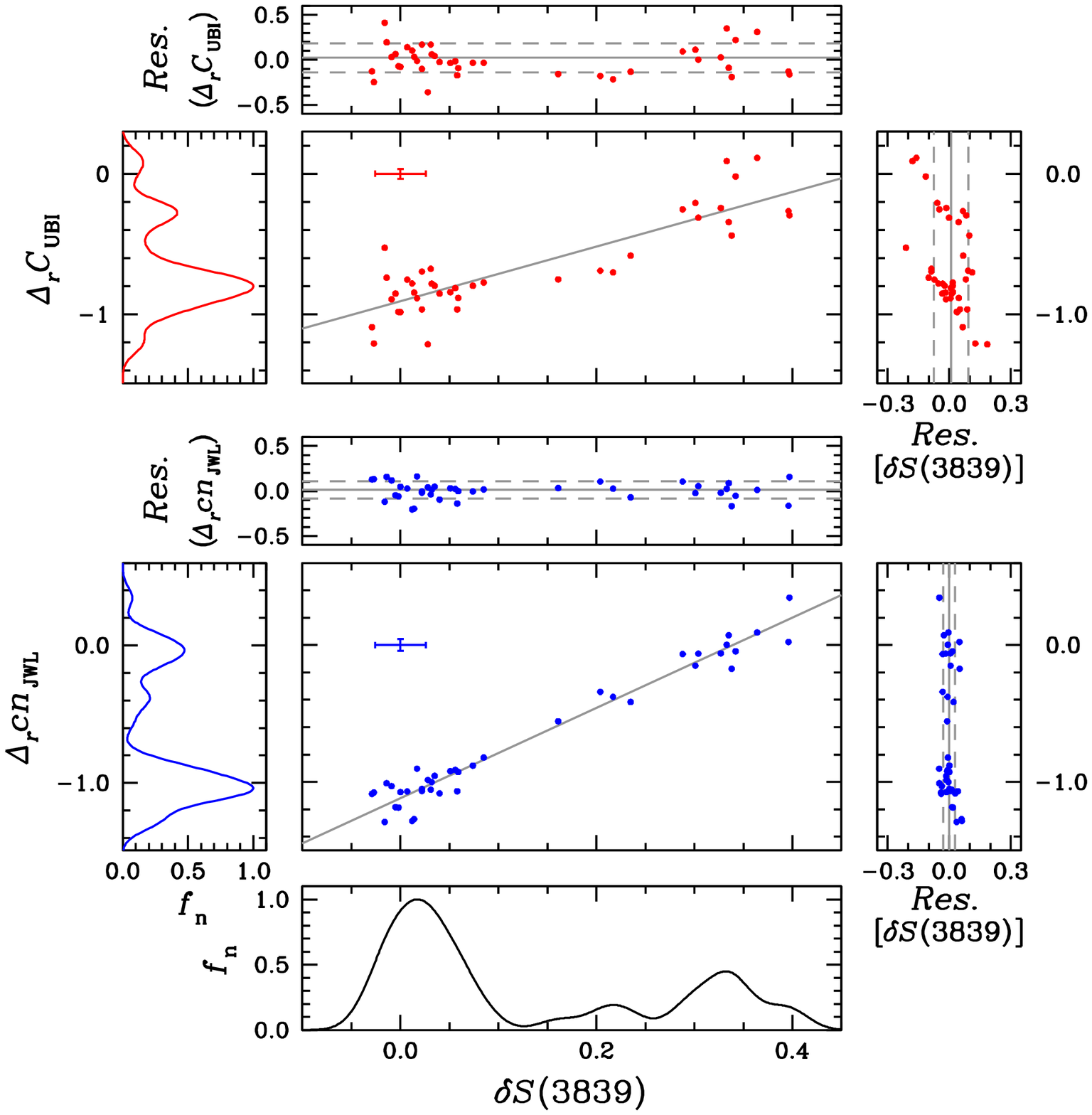}
\caption{
Plots of \ds\ versus \cnjwl\ and \dcubi\ for M3 RGB stars along with least--square fits. The mean residuals ($\pm 1\sigma$) around the fitted lines are also shown with long-dashed lines. The \dcubi\ index shows a weak correlation with the spectroscopic \ds\ index by \citet{smolinski11}. In sharp contrast, the plot shows that our \dcnjwl\ index is nicely correlated with the \ds, with the correlation coefficient of $\rho$ = 0.981 (see Table~\ref{tab:fit}). We emphasize that the shape of the histogram for \dcnjwl\ is very similar to that for the \ds. Also note that the scatter around the fitted lines are $\sigma$ = 0.163 and 0.094 for the \dcubi\ and \dcnjwl, respectively, and they are at the levels of 4.5$\times\sigma$(\dcubi) and 2.2$\times\sigma$(\dcnjwl), where $\sigma$(\dcubi) and $\sigma$(\dcnjwl) are photometric measurement uncertainties. In the right panels, we show the residuals in \ds\ around the fitted lines, finding $\sigma$ = 0.084 and 0.029 for the \dcubi\ and \dcnjwl, respectively, and they are at the levels of 3.2$\times\sigma$[\ds] for \dcubi\ and 1.1$\times\sigma$[\ds] for \dcnjwl, where $\sigma$[\ds] is the spectroscopic measurement uncertainty by \citet{smolinski11}.
The large scatter in \dcubi\ is not due to the photometric or the spectroscopic 
measurement errors but due to its intrinsic nature as a poor CN--tracer.
}\label{fig:m3}
\end{figure}
%=======================================================================

%=======================================================================
\begin{deluxetable}{lccccc}
\tablenum{3}
\tablecaption{The Goodness of the Fit \label{tab:fit}}
\tablewidth{0pc}
\tablehead{
\multicolumn{1}{c}{} &
\multicolumn{2}{c}{\dcnjwl} &
\multicolumn{1}{c}{} &
\multicolumn{2}{c}{\dcubi}\\
\cline{2-3}\cline{5-6}
\multicolumn{1}{c}{} &
\multicolumn{1}{c}{$p$-value} &
\multicolumn{1}{c}{$\rho$} &
\multicolumn{1}{c}{} &
\multicolumn{1}{c}{$p$-value} &
\multicolumn{1}{c}{$\rho$} 
}
\startdata
\ds    & 0.000 &    0.981 && 0.000 & 0.862 \\
\enddata 
\end{deluxetable}
%=======================================================================

\section{POPULATIONAL TAGGING}\label{s:tag}

\subsection{Comparison with \citet{smolinski11}}
In our previous work for M5 and NGC~6752, we showed that our \cnjwl\ index accurately traces the CN contents of the cool giant stars in GCs \citep{lee17,lee18}. Here we explore the nature of the \dcubi\ index as a population tagger.

\citet{smolinski11} presented the homogeneous CN and CH absorption band strengths for stars in eight GCs, including M3, obtained during the course of the Sloan Extension for Galactic Understanding and Exploration subsurvey of the SDSS. In Figure~\ref{fig:m3}, we show comparisons of the \ds\ versus \dcnjwl\ and \dcubi.

We calculated the Pearson's correlation coefficient for the \cnjwl\ versus \ds\ and we obtained $\rho$ = 0.981 with a $p$-value of 0.000 (see also Table~\ref{tab:fit}). As can be seen, the scatters in the \dcnjwl\ around the fitted line are very small. The mean residual in the \dcnjwl\ is $\sigma$ = 0.094, which is at the level of 2.2$\times\sigma$(\dcnjwl), where $\sigma$(\dcnjwl) is the mean photometric measurement uncertainty of the RGB stars used in the figure. On the other hand, the residual in \ds\ around the fitted line is $\sigma$ = 0.029 and it is at the level of 1.1$\times\sigma$[\ds], where $\sigma$[\ds] is the mean spectroscopic measurement uncertainty by \citet{smolinski11}. It should be noted that the stars studied by \citet{smolinski11} are rather isolated ones and, as a consequence, the photometric measurements errors are not expected to be large. Therefore, the small scatter in the correlation between \dcnjwl\ and \ds\ can be fully explained by the spectroscopic measurement uncertainties reported by \citet{smolinski11}. We also note that the shapes of the histogram for both the \ds\ and \dcnjwl\ are almost identical, except for the scale.

From a statistical point of view, the correlation between the \dcubi\ and the \ds\ can be considered to be decent, with a Pearson's correlation coefficient of $\rho$ = 0.862
with a $p$-value of 0.000. However, the scatter around the fitted line is about twice as large as that from the \dcnjwl, $\sigma$ = 0.163, which is at the level of 4.5$\times\sigma$(\dcubi). Also, the residual in \ds\ around the fitted line is quite large, $\sigma$ = 0.084, which is at the level of 3.2$\times\sigma$[\ds]. It is naturally thought that the large scatter in the correlation between \dcubi\ and \ds\ is not due to photometric or spectroscopic measurement uncertainties but due to its poor nature as a CN tracer. Contrary to the \dcnjwl\ index, the substructure of the histogram of the \dcubi\ is slightly different from that of the \ds. Also, it should be pointed out that a hint of a nonlinear correlation between the \dcubi\ and the \ds\ can be seen in the figure.

From this exercise, we can conclude that our \dcnjwl\ index is precisely correlated with the spectroscopic \ds\ index, while the \dcubi\ index is not and should be used with caution.

%=======================================================================
\begin{deluxetable*}{lccc}
\tablenum{4}
\tablecaption{Fractions of the FG population (\cnw\ or \dcubi-blue) 
within $r \leq 91\arcsec$\label{tab:2milone}}
\tablewidth{0pc}
\tablehead{
\multicolumn{1}{c}{Name} &
\multicolumn{1}{c}{\dcnjwl} &
\multicolumn{1}{c}{\dcubi} &
\multicolumn{1}{c}{\citet{milone17}}
}
\startdata
NGC~5272 (M3) & 0.332 ($\pm$ 0.040) & 0.475 ($\pm$ 0.050) & 0.305 ($\pm$ 0.014) \\
NGC~5904 (M5) & 0.320 ($\pm$ 0.023) & 0.328 ($\pm$ 0.025) & 0.235 ($\pm$ 0.013) \\
NGC~6752      & 0.280 ($\pm$ 0.039) & 0.185 ($\pm$ 0.032) & 0.294 ($\pm$ 0.023) \\ 
\enddata 
\end{deluxetable*}
%=======================================================================

%=======================================================================
\begin{figure}
\epsscale{1.2}
\figurenum{5}
\plotone{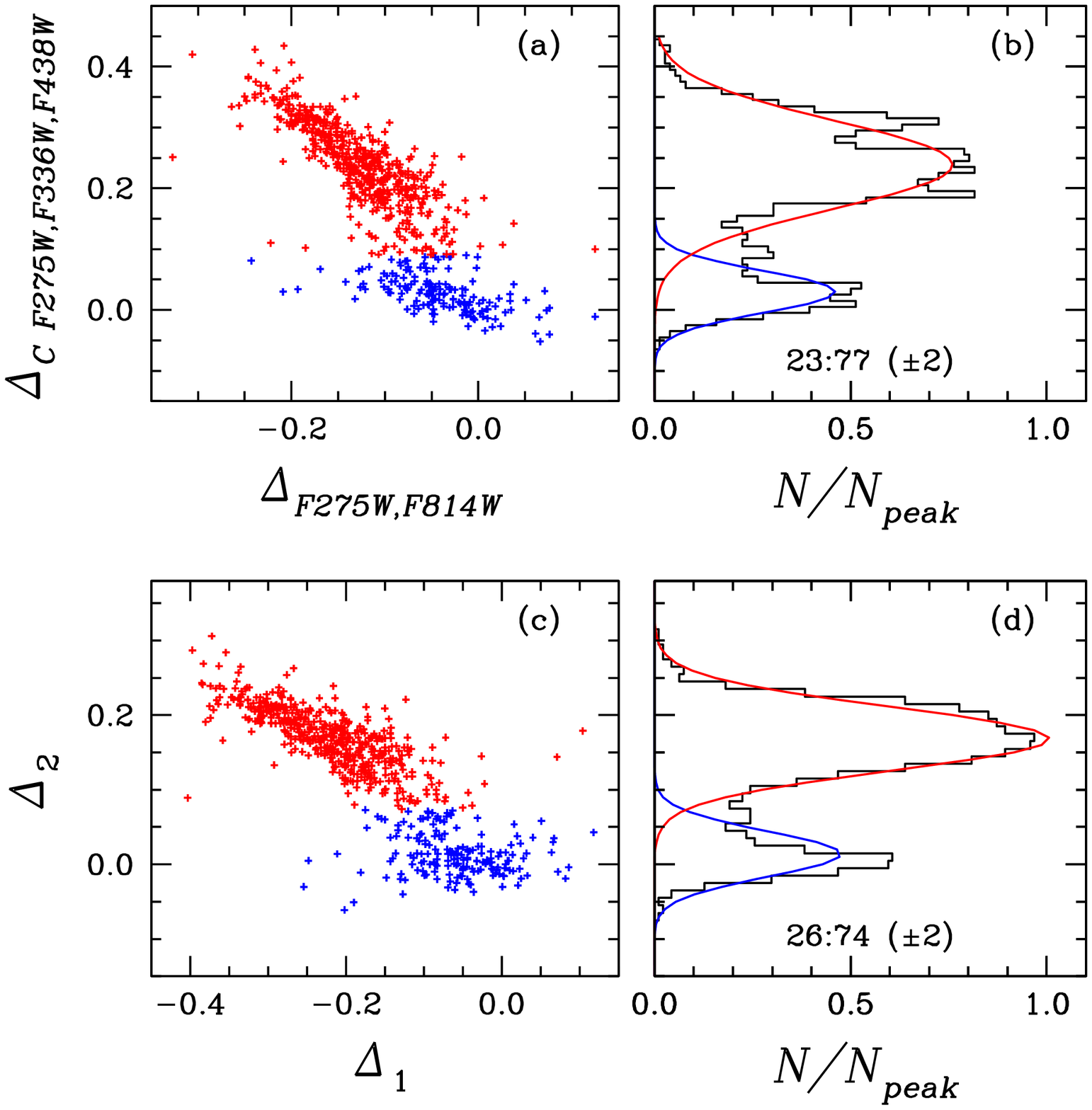}
\caption{(a) Plot of \dc\ versus \dtrio\ of M5 RGB stars with $-2 \leq$ \vvhb\ $\leq$ 2 mag using the \hst\ photometry by \citet{milone17}. The blue and red plus signs denote the FG and SG of the cluster from our expectation maximization estimator. 
(b) Histogram of the \dtrio\ distribution with the populational number ratio of \nfgsg\ = 23:77 ($\pm$ 2).
(c) Same as panel (a), but for $\Delta_1$ versus $\Delta_2$. 
(d) Histogram of the $\Delta_2$ distribution with the populational number ratio of \nfgsg\ = 26:74 ($\pm$ 2).
}\label{fig:m5hst}
\end{figure}
%=======================================================================

\subsection{Comparison with \citet{milone17}}
We compared our populational number ratios of the central part of individual clusters with those of \citet{milone17}. Note that the FOV of the \hst\ WFC3/UVIS (= $162\arcsec\times 162\arcsec$) is significantly smaller than that of our observations ( $> 30\arcmin\times 30\arcmin$). Therefore, in our calculations of the populational number ratios, we restricted stars within the radial distance of 91\arcsec\ in our results to maintain the same projected area on the sky between our work and \citet{milone17}.

The very dense environment in the central part of GCs could induce a potential problem in the populational tagging, especially based on any ground-based photometry \citep[see the Appendix of][]{lee17}. In our ALLFRAME run, the detection of stars was performed using our mosaicked master frames, which were constructed with a sufficiently large number ($>$ 150) of science frames from various passbands with good seeing conditions. The differences in the color indices, for example, ($b-y$) and \cnjwl, between the two populations are negligibly small; therefore, at the given magnitude and the degree of crowdedness, there would exist no color effect in detecting stars from different populations. However, GCs with strong radial gradient in the populational number ratio could be a problematic, since the more centrally concentrated population has a higher probability of not detecting in the central part of the cluster, and as a consequence, an incorrect populational number ratio can be inferred. As we will show later, M3 is such a cluster showing a very strong radial gradient in the populational number ratio. The populational number ratio in the central part of M3 from our ground-based observations is in excellent agreement with that from the \hst\ WFC3/UVIS, and the radial populational gradient does not affect our results presented here.

We calculated the populational number ratios using the expectation maximization (EM) method with the two-component gaussian mixture model, assuming two stellar populations for each cluster \citep[see also][]{lee15,lee17,lee18}. In an iterative manner, we calculated the probability of individual stars for being the \cnw\ and \cns\ populations. Stars with $P$(\cnw$|x_i) \geq$ 0.5 from the EM estimator are corresponding to the \cnw\ population, where $x_i$ are the individual RGB stars, while those with $P$(\cns$|x_i)$ $>$ 0.5  represent the \cns\ population. Through this process, we securely obtained the populational number ratio between the two different stellar populations.

In Table~\ref{tab:2milone}, we show our results. For M3 and NGC~6752, the populational number ratios from our \dcnjwl\ index in the central parts of the clusters are in excellent agreement with those from \citet{milone17}, who calculated the fractions of the FG in each cluster based on the distribution of the $\Delta_2$ index, which will be discussed below.

For M5, the difference in the populational number ratio between our result and that of \citet{milone17} is rather large (at the level of 2.4$\sigma$) compared to M3 and NGC~6752. As we will show later, M5 does not show any strong radial gradient in the populational number ratio, and the origin of this discrepancy is not clearly understood. It should be reminded that our \dcnjwl\ index and the $\Delta_2$ index devised by \citet{milone17} have completely different bases. Unlike our \dcnjwl\ index, which is specifically designed to measure the CN band at \cnwave\ \AA\ only, the $\Delta_2$ (or \trio) index measures very strong absorption features, including OH, NH, CN, and CH. As \citet{milone15} nicely demonstrated, the $\Delta_2$ index is expected to utilize the different elemental abundance dependencies on the various filters to maximize the separation  between the multiple stellar populations. It is difficult to believe that crowdedness in the central part of the cluster is responsible for the discrepancy in the number ratios from \cnjwlcor\ and $\Delta_2$. As we elaborately showed in our previous work, the crowdedness does not affect the populational number ratio in M5 \citep[see, e.g., the appendices of][]{lee17}.

The magnitude ranges of the RGB stars used by \citet{milone17} are different from that of our work, $-2 \leq$ \vvhb\ $\leq$ 2 mag, where \vhb\ is the visual magnitude of the horizontal branch (HB) stars of the cluster. Both in our \dcnjwl\ and the \trio\ indices, the level of confusion in the populational separation in the lower RGB stars would become greater. Therefore, we reanalyzed the populational number ratios for M5  RGB stars with $-2 \leq$ \vvhb\ $\leq$ 2 mag using the \hst\ photometry by \citet{milone17}. In Figure~\ref{fig:m5hst}, we show our results. In the figure, we used the following relations to calculate $\Delta_1$ and $\Delta_2$;
\begin{eqnarray}
\Delta_2 & = & X\times\sin{\theta} + Y\times\cos{\theta}, \\
\Delta_1 & = & X\times\cos{\theta} - Y\times\sin{\theta}, 
\end{eqnarray}
where $X$ = \dc\ and $Y$ = \dtrio \footnote{See \citet{milone17}  for the definitions of the \dc\ and the \dtrio\ indices.}. In our calculations, we adopted $\theta$ = 25$\arcdeg$ for M5. By applying the EM estimator with the two-component gaussian mixture model, we obtained the fraction of the FG of 0.260 $\pm$ 0.022, which is slightly larger than that by \citet{milone17}. If we adopt our new value, the discrepancy in the populational number ratio for M5 becomes smaller, and the results from our \dcnjwl\ index and our $\Delta_2$ are in agreement to within a 1.3$\sigma$ level. From our demonstrations, we can conclude that the results from our \dcnjwl\ index are consistent with those from \hst\ UV photometry.

In Table~\ref{tab:2milone}, we also show the results for \dcubi. The fractions of \dcubi-blue are significantly different from those of \citet{milone17}, suggesting that the \cubi\ index may not be reliable.

We made more extensive comparisons between our \dcnjwl\ and the \dcubi\ indices to explore the utility of the \cubi\ index in the field of MPs of GCs, and we show our results below.

%=======================================================================
\begin{deluxetable}{lcc}
\tablenum{5}
\tablecaption{Populational Number Ratios \label{tab:pop}}
\tablewidth{0pc}
\tablehead{
\multicolumn{1}{c}{Name} &
\multicolumn{1}{c}{\dcnjwl} &
\multicolumn{1}{c}{\dcubi} }
\startdata
NGC~5272 (M3) & 48:52 ($\pm$ 3) & 57:43 ($\pm$ 3)  \\
NGC~5904 (M5) & 30:70 ($\pm$ 2) & 34:66 ($\pm$ 3)  \\
NGC~6752      & 27:73 ($\pm$ 3) & 25:75 ($\pm$ 3)  \\ 
\enddata 
\end{deluxetable}
%=======================================================================

%=======================================================================
\begin{figure}
\epsscale{1.2}
\figurenum{6}
\plotone{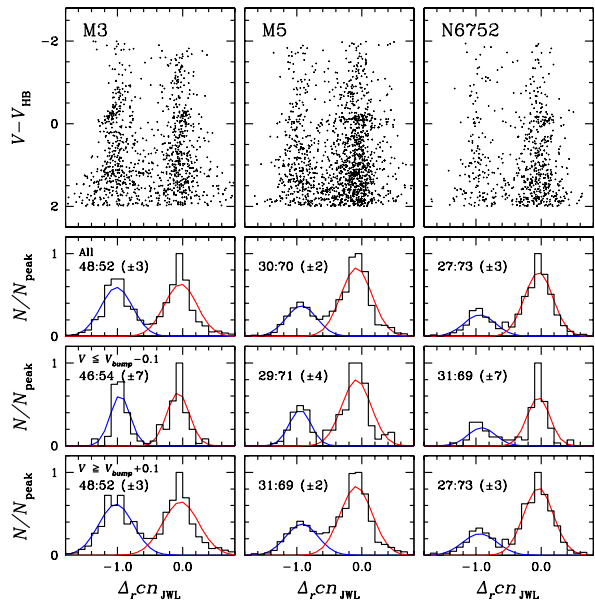}
\caption{
The \dcnjwl\ CMDs of RGB stars with $-$2 $\leq$ \vvhb\ $\leq$ 2 mag and the \dcnjwl\ distributions for all RGB stars, the RGB stars brighter than the RGB bump, and the RGB stars fainter than the RGB bump. The populational number ratios between the two magnitude levels are in excellent agreement within the statistical fluctuations, indicative of the absence of the internal mixing effect in these three GCs.
}\label{fig:str}
\end{figure}
%=======================================================================

%=======================================================================
\begin{figure}
\epsscale{1.2}
\figurenum{7}
\plotone{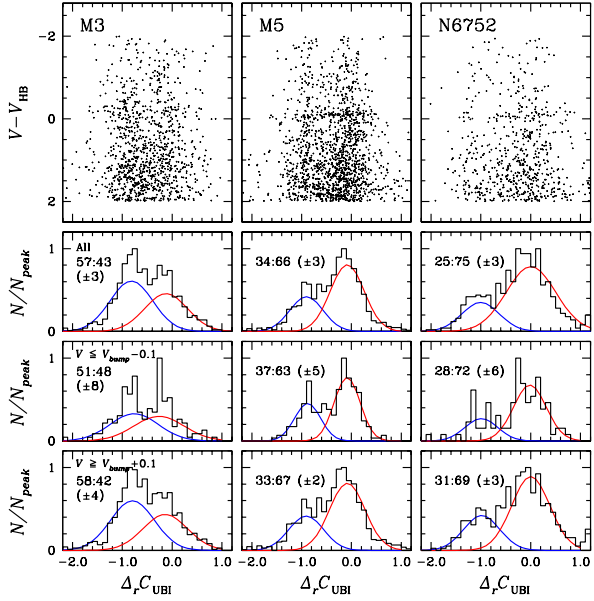}
\caption{
Same as Figure~\ref{fig:str}, but for \dcubi. The separation between the assumed two populations are not clear, implying that the \dcubi\ index suffers from severe confusion in populational tagging.
}\label{fig:ubi}
\end{figure}
%=======================================================================

\subsection{Populational number ratio}
In Figure~\ref{fig:str}, we show the CMDs of RGB stars around the visual magnitude of the HB stars and populational number ratios from our \dcnjwl\ index. As can be seen, each histogram for our \dcnjwl\ distribution shows the conspicuous double peaks with the clear populational separation. Using the EM estimator with two-Gaussian mixture model, we obtained the number ratios between the two populations of \nrgb\ = 48:52 ($\pm$ 3), 29:71 ($\pm$ 2), and 25:75 ($\pm$ 3) for M3, M5, and NGC~6752, respectively. Note that the populational number ratios for M5 and NGC~6752 are the same as those without the rectification process that we calculated in our previous works \citep{lee17,lee18}, implying that our rectification processes using the equations (\ref{eq1}) and (\ref{eq2}) do not distort the populational number ratios of the clusters. Also shown in the figure, the populational number ratios from the two different magnitude levels are in excellent agreement, indicative of the absence of the internal mixing effect on the populational number ratios in these three clusters \citep[see, e.g.,][]{lee10}.

Figure~\ref{fig:ubi} shows the populational number ratios from the \dcubi\ index. We obtained the number ratios of $n$(\dcubi-blue):$n$(\dcubi-red) = 57:43 ($\pm$ 3), 34:66 ($\pm$ 3) and 25:75 ($\pm$ 3) for M3, M5, and NGC~6752, respectively, and we summarized our results in Table~\ref{tab:pop}. The populational number ratios for M5 and NGC~6752 from the \dcubi\ index are in agreement with those from the \dcnjwl\ within the statistical errors, while the populational number ratios for M3 from the both indices do not agree within the measurement errors. In the figure, it should be emphasized that the transition from one population  to the other is not clear in \dcubi, and therefore the populational tagging for individual stars can be somewhat uncertain.

%=======================================================================
\begin{figure}
\epsscale{1.2}
\figurenum{8}
\plotone{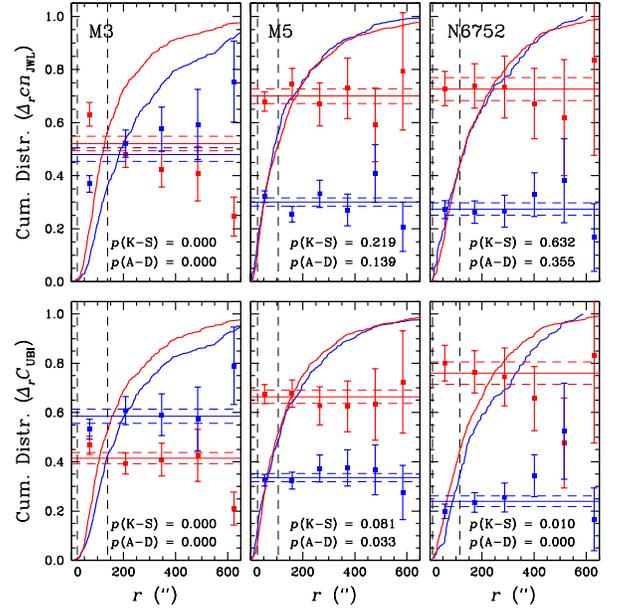}
\caption{Top panels: Comparisons of the cumulative radial distributions and fractions of the \cnw\ (blue solid lines) and  \cns\ (red solid lines). The mean values and the standard deviations for the fractions of the \cnw\ and \cns\ populations are given with the horizontal solid and long-dashed lines, respectively, while the fractions of each population in each half-light radii are given with the filled square with error bars. The gray long-dashed lines denote the core ($r_c$) and half-light ($r_h$) radii.
Bottom panels: Same as the top panels, but for the \dcubi-blue and \dcubi-red populations.
}\label{fig:rad}
\end{figure}
%=======================================================================

%=======================================================================
\begin{deluxetable}{lccccc}
\tablenum{6}
\tablecaption{The $p$-values returned from the K--S and A--D tests for cumulative radial distributions\label{tab:ks}}
\tablewidth{0pc}
\tablehead{
\multicolumn{1}{c}{Name} &
\multicolumn{2}{c}{\dcnjwl} &
\multicolumn{1}{c}{} &
\multicolumn{2}{c}{\dcubi} \\
\cline{2-3}\cline{5-6}
\multicolumn{1}{c}{} &
\multicolumn{1}{c}{K--S} &
\multicolumn{1}{c}{A--D} &
\multicolumn{1}{c}{} &
\multicolumn{1}{c}{K--S} &
\multicolumn{1}{c}{A--D} }
\startdata
NGC~5272 (M3) & 0.000 & 0.000 && 0.000 & 0.000  \\
NGC~5904 (M5) & 0.219 & 0.139 && 0.081 & 0.033  \\
NGC~6752      & 0.632 & 0.355 && 0.010 & 0.000  \\
\enddata 
\end{deluxetable}
%=======================================================================

\subsection{Cumulative radial distribution}
The cumulative radial distributions of the MPs in GCs can provide some fundamental information on the dynamical evolution of GCs, although the time scale required for the complete homogenization does not appear to be clear \citep[see, e.g.,][and references therein]{lee17,lee18}.

In Figure~\ref{fig:rad}, we show comparisons of the cumulative radial distributions for each cluster. As shown in the figure, our results strongly suggest that the cumulative radial distributions of individual populations can be different between those from the \dcnjwl\ and the \dcubi\ indices, in spite of similar populational number ratios between the two approaches. We performed Kolmogorov--Smirnov (K--S) tests to examine whether the two populations of individual clusters from each color index are identical. In Table~\ref{tab:ks}, we show the $p$-values for the null hypothesis that the two populations are drawn from identical parent distributions. For the K--S tests of our \dcnjwl\ index, the two populations in M3 are not drawn from  identical parent distributions, while those in M5 and NGC~6752 are likely drawn from identical parent distributions \citep[see also][]{lee17,lee18}.

On the other hand, a very different conclusion can be drawn from the K-S tests of the \dcubi\ index for NGC~6752 and, perhaps, for M5: from the \dcubi\ point of view, the two populations in NGC~6752 are not drawn from identical parent distributions with a $p$-value of 0.010. The $p$-value of 0.081 for M5 is thought to be on the verge of either rejecting or accepting the null hypothesis.

It is a well-known fact that the K-S test can be sensitively dependent on the near center (or the median) of the distribution and less dependent on the edges of the distribution. We performed the $k$-sample Anderson--Darling (A--D) tests, which are known to be less vulnerable to such problem, and we also show our results in Table~\ref{tab:ks}. From our A-D tests, it becomes very clear that  the cumulative radial distributions from \dcubi\ are very different from those from \dcnjwl\ for M5 and NGC~6752: the two stellar populations classified in \dcubi\ in M5 are not most likely drawn from the identical parent distribution.

For M3, the $p$-values from both the K-S and the A-D tests indicate that the cumulative radial distributions of the two MPs are definitely different. However, it should be emphasized in Figure~\ref{fig:rad} that the central concentration  of the \cns\ population from the \dcnjwl\ index is larger than that of the \dcubi-red population from the \dcubi\ index for M3. Also, a strong radial gradient in the population number ratios can be seen in the distributions from the \dcnjwl, while a weak (or a flat) gradient can be seen in those from the \dcubi.

As we already discussed in \S4.2, it is very interesting to note that \citet{milone17} found a fraction of the first generation of stars of 0.305 $\pm$ 0.014 based on the \hst\ observations. Assuming that their definition of the first generation of the stars is the same as our definition for the \cnw\ population, which we already confirmed for M5 and NGC~6752 \citep{lee17,lee18}, their estimation does not agree with our mean fraction of the \cnw\ population of the cluster from a large FOV ($r \lesssim 5r_h$), 0.479 $\pm$ 0.029. The figure shows that the \cns\ population of M3 is more centrally concentrated and shows a very strong radial gradient. As we discussed before, the fraction of the \cnw\ population from our \dcnjwl\ index within the same FOV of the \hst\ WFC3/UVIS is 0.332 $\pm$ 0.040 and it is in excellent agreement with the result reported by \citet{milone17}. Therefore, the discrepancy in the fraction of the first generation of stars between our study and \citet{milone17} for M3 highlights that the statistical study of the MPs in GCs from a large FOV is essential.

The result from \citet{smolinski11} also nicely demonstrated the opposite situation. They found that the number ratio between the CN-weak and CN-strong star in M3 is 56:44  (see their Figure~4 and Table~4). Due to the crowdedness of the central part of the cluster, the spectroscopic target stars used by \citet{smolinski11} are restricted in the outer part of the cluster. Due to the central concentration and the strong radial gradient of the populational ratio, it is expected that the number ratio for the CN-weak stars will become greater in the outer part of the cluster, consistent with the result by \citet{smolinski11}. Our exercises vividly demonstrate why we need to rely on the \cnjwl\ index in order to perform a precision study of the MPs in the central part of the clusters.

For NGC~6752, the populational fractions remain flat or show a weak radial gradient in \dcnjwl, while a strong radial gradient can be seen in \dcubi, in spite of the identical mean populational fractions between the two methods.

\section{SUMMARY AND CONCLUSION}
It is believed that the majority of the monometallic normal GCs in the Milky Way Galaxy contain MPs. The variations in abundances of some elements, which are not synthesized or destroyed during the course of the early evolution of the low-mass stars, can be smoking-gun evidence of the chemical pollution by the previous generations of the stars. In particular, nitrogen and carbon show drastic elemental abundance variations between the MPs in GCs, and these species formed diatomic molecules that significantly change the visible part of the spectra in the RGB stars through NH, CN, and CH, whose absorption strengths are so strong that even broadband photometry can be affected.

As we already showed in our previous work, the $m1$ and the $cy$ from the \str\ photometry and the \dug\ index from the SDSS photometry can only provide some limited information on the degree of the variations of the lighter elemental abundances \citep{lee17,lee18}. The classifications of the MPs from these color indices are seemingly similar, but the influences from the various elements on these color indices hinder precision populational tagging, which is the fundamental basis of the study of the MPs in GCs, such as the populational number ratios, the cumulative radial distributions, spatial distributions, etc.

The utilization the widely used $UBI$ photometry is undeniably attractive, since there already exist a huge amount of archival data for the most of the GCs in our Galaxy. Here we performed the critical assessment of the utility of \cubi\ for precision populational tagging. Our study clearly showed that the \cubi\ index has an inherited trait of the broad-band photometry: a confusion due to its dependency on the multiple elements, as we already showed for $m1$, $cy$, and \dug, for example. In particular, the populational tagging for the RGB stars fainter than the HB magnitude level becomes very ambiguous in the \cubi\ index, and as a consequence, the populational number ratios and  the cumulative radial distributions derived from the \cubi\ index may not be trustworthy. Our results strongly suggested that the \cubi\ index is dependent on various elemental abundances and stellar parameters. The decompositions of various effects on the \cubi\ index could be a formidable or even an impossible task in the framework of the broadband $UBI$ photometry.

In sharp contrast, our \cnjwl\ index is really a measure of the CN strength, which is known to be governed by the nitrogen abundance, of the cool stars in GCs. We strongly believed that our \cnjwl\ index can perform the most precise population tagging in the central part of the GCs, where  the spectroscopic approach can not be applied due to crowdedness.

The discrepancy between our populational number ratio and that by \citet{smolinski11} for M3 highlights the strengths of our approach. \citet{smolinski11} measured the populational number ratio of \nrgb\ = 56:44 for M3, which is slightly different from our result, 48:52. M3 has a very strong radial gradient in the populational number ratio, in the sense that the \cns\ population is more centrally concentrated. Since the spectroscopic measurements by \citet{smolinski11} tended to be based on the outer part of M3, it is natural to mistakenly have a lower fraction of the \cns\ population in the outer part of the cluster.

Similarly, it can be understood that a significantly low fraction of the FG population in the central part of M3 by \citet{milone17} is due to the presence of the strong radial gradient in the populational number ratio in M3. This also highlights the importance of the securing a large FOV to perform a precision MP study of the clusters.

Finally, comparisons of the populational number ratios between the RGB stars fainter and brighter than the RGB bump visual magnitudes implied that these three GCs do not show any differences in the number ratios, indicative of the absence of the internal mixing effect on the lighter elemental abundances \citep[see, e.g.,][]{lee10}.

\acknowledgements
J.-W.L. acknowledges financial support from the Basic Science Research Program (grant no. 2016-R1A2B4014741) through the National Research Foundation of Korea (NRF) funded by the Korea government (MSIP). He also tanks Dr. Stetson for providing $UBI$ photometry for M5 and NGC~6752 used in this study and the anonymous referee for constructive comments.

\end{document}